
\font\lbf=cmbx10 scaled\magstep2

\def\bs{\bigskip}
\def\ms{\medskip}
\def\np{\vfill\eject}

\def\ni{\noindent}
\def\cl{\centerline}

\def\title#1{\cl{\lbf #1}}
\def\ref#1#2#3#4{#1\ {\it#2\ }{\bf#3\ }#4\par}
\def\refb#1#2#3{#1\ {\it#2\ }#3\par}
\def\ANY{Ann.\ N.Y.\ Acad.\ Sci.}
\def\CQG{Class.\ Qu.\ Grav.}
\def\CMP{Comm.\ Math.\ Phys.}
\def\CPAM{Comm.\ Pure App.\ Math.}
\def\JMP{J.\ Math.\ Phys.}
\def\PL{Phys.\ Lett.}
\def\PR{Phys.\ Rev.}
\def\PRL{Phys.\ Rev.\ Lett.}
\def\PRS{Proc.\ Roy.\ Soc.\ Lond.}
\def\PTP{Prog.\ Theor.\ Phys.}

\def\sq{\rlap{$\sqcup$}$\sqcap$}
\def\l{\lim_{\gamma\to p}}
\def\r{\lim_{r\to\infty}}
\def\d{\hbox{d}}
\def\e{\hbox{e}}
\def\p{\partial}
\def\I{\int_\Sigma{*}}
\def\O#1{\left.#1\right|_{\p\Sigma}}
\def\R{{\cal R}}
\def\T{{\cal T}}
\def\E{{\cal E}}
\def\M{{\cal M}}
\def\scri{{\cal I}}
\def\half{{\textstyle{1\over2}}}
\def\quart{{\textstyle{1\over4}}}
\def\ft{{\textstyle{4\over3}}}

\magnification=\magstep1

\title{Gravitational energy in spherical symmetry}
\bs\cl{\bf Sean A. Hayward}
\ms\cl{Department of Physics, Kyoto University, Kyoto 606-01, Japan}
\ms\cl{\tt hayward@murasaki.scphys.kyoto-u.ac.jp}
\bs\ni{\bf Abstract.}
Various properties of
the Misner-Sharp spherically symmetric gravitational energy $E$
are established or reviewed.
In the Newtonian limit of a perfect fluid,
$E$ yields the Newtonian mass to leading order
and the Newtonian kinetic and potential energy to the next order.
For test particles, the corresponding H\'aj\'\i\v cek energy is conserved
and has the behaviour appropriate to energy
in the Newtonian and special-relativistic limits.
In the small-sphere limit, the leading term in $E$
is the product of volume and the energy density of the matter.
In vacuo, $E$ reduces to the Schwarzschild energy.
At null and spatial infinity,
$E$ reduces to the Bondi-Sachs and Arnowitt-Deser-Misner energies respectively.
The conserved Kodama current has charge $E$.
A sphere is trapped if $E>\half r$,
marginal if $E=\half r$ and untrapped if $E<\half r$,
where $r$ is the areal radius.
A central singularity is spatial and trapped if $E>0$,
and temporal and untrapped if $E<0$.
On an untrapped sphere,
$E$ is non-decreasing in any outgoing spatial or null direction,
assuming the dominant energy condition.
It follows that $E\ge0$ on an untrapped spatial hypersurface
with regular centre,
and $E\ge\half r_0$ on an untrapped spatial hypersurface
bounded at the inward end by a marginal sphere of radius $r_0$.
All these inequalities extend to the asymptotic energies,
recovering the Bondi-Sachs energy loss
and the positivity of the asymptotic energies,
as well as proving the conjectured Penrose inequality for black or white holes.
Implications for the cosmic censorship hypothesis
and for general definitions of gravitational energy are discussed.
\bs\cl{PACS: 04.70.Bw, 04.20.Dw, 04.20.Ha, 04.25.Nx}
\bs\cl{Revised 6th November 1995}
\np\ni
{\bf I. Introduction}
\ms\ni
A massive source produces a gravitational field which has energy.
In Relativity theory, the equivalence of mass and energy means that
it is only the combined energy which may be measured at a distance.
Moreover, the non-linearity of the gravitational field means that
the material (or passive) mass and its gravitational and kinetic energy
combine in a non-linear, non-local way
to produce the effective (or active) energy.
In spherical symmetry, in vacuo,
this effective energy is just the Schwarzschild energy.
In general, there is no agreed definition of this energy,
except at infinity in an asymptotically flat space-time,
where one has
the Arnowitt-Deser-Misner [1] energy $E_{ADM}$ at spatial infinity
and the Bondi-Sachs [2--3] energy $E_{BS}$ at null infinity.
One would therefore like a definition of energy
which reduces to these asymptotic energies appropriately.
Also, given the above physical motivation,
one would like the energy to yield the Newtonian mass in the Newtonian limit,
with the highest-order correction yielding the Newtonian energy.
Similarly, one would like the energy to yield the correct energies
of test particles in the Newtonian and special-relativistic limits.
Remarkably, such an energy does exist in spherical symmetry:
the Misner-Sharp [4] energy $E$.
Moreover, $E$ is intimately related to
the characteristic strong-field gravitational phenomena,
namely black and white holes and singularities.

This article lists various key properties of this energy.
The main new results,
apart from the Newtonian, test-particle and special-relativistic limits,
are a monotonicity property of $E$ which leads to a positivity property
and a lower bound for $E$ in terms of the area of a black or white hole,
the so-called isoperimetric inequality [5].
Here, black and white holes are defined by marginal surfaces of certain types,
as explained later, or in [6] or [7].
These properties of $E$ extend to the asymptotic energies,
in particular establishing the isoperimetric inequality for $E_{ADM}$,
as conjectured by Penrose [5], and for $E_{BS}$.
Known results are also reviewed,
partly because some are prerequisites for the newer results,
and partly because the existing literature on the subject
is somewhat dispersed,
so that it is not always appreciated that
$E$ enjoys quite so many desirable properties.

In Section II, $E$ is defined geometrically
and shown to have various purely geometrical properties
related to trapped and marginal surfaces, central singularities
and the asymptotic energies.
In Section III, various dynamical properties are derived
assuming the dominant energy condition,
including the monotonicity, positivity and area-bound properties.
A discussion of implications for the cosmic censorship hypothesis
is also given there.
In Section IV, the geometry is decomposed with respect to spatial hypersurfaces
and the behaviour of $E$ in the Newtonian limit is found.
In Section V, the energy of test particles as determined by $E$ is discussed,
including the Newtonian and special-relativistic limits.
The Conclusion discusses
the implications for more general suggestions for gravitational energy.
The Appendices concern (A) Newtonian theory,
(B) the identity of $E$ as a charge associated with a conserved current,
and (C) energy-momentum.

Since the whole article is concerned with spherical symmetry,
this case will be assumed implicitly in the Propositions,
without repeated qualification.
Similarly, all geometrical objects mentioned
will be assumed to respect the spherical symmetry.
All arguments will be local,
except to note implications for conformal infinity, if it exists.
\np\ni
{\bf II. Geometrical properties}
\ms\ni
The line-element may be written locally in double-null form as
$$\d s^2=r^2\d\Omega^2-2\e^{-f}\d\xi^+\d\xi^-\eqno(1)$$
where $\d\Omega^2$ refers to the unit sphere
and $r$ and $f$ are functions of the null coordinates $(\xi^+,\xi^-)$.
This double-null form is natural in the sense that
each symmetric sphere has two preferred normal directions,
the null directions $\p/\p\xi^\pm$.
One may also use one spatial and one temporal direction, as in Section IV,
but there is no unique choice of such directions,
which makes it more difficult to check coordinate invariance.
In double-null form,
the remaining coordinate freedom consists simply of the diffeomorphisms
$$\xi^\pm\mapsto\hat\xi^\pm(\xi^\pm).\eqno(2)$$
The geometry is given by the metric (1) modulo the null rescalings (2).
Note that $r$ is a geometrical invariant but that $f$ is not.
The area of a symmetric sphere is $4\pi r^2$,
so that $r$ is the areal radius, and will simply be called the {\it radius}.
One may take $r>0$, with $r=0$ being discussed shortly.

The space-time will be assumed time-orientable,
and $\p/\p\xi^\pm$ will be assumed future-pointing.
The {\it expansions} may be defined by
$$\theta_\pm=2r^{-1}\p_\pm r\eqno(3)$$
where $\p_\pm$ denotes the coordinate derivative along $\xi^\pm$.
The expansions measure whether the light rays normal to a sphere
are diverging ($\theta>0$) or converging ($\theta<0$),
or equivalently, whether the area of the spheres
is increasing or decreasing in the null directions.
Note that the signs of $\theta_\pm$ are geometrical invariants,
but their actual values are not.
An invariant combination is $\e^f\theta_+\theta_-$,
or equivalently $g^{-1}(\d r,\d r)=-\half\e^fr^2\theta_+\theta_-$,
where $g$ is the space-time metric.
Indeed, the only invariants of the metric and its first derivatives
are functions of $r$ and $\e^f\theta_+\theta_-$.
The latter invariant has an important geometrical and physical meaning:
a metric sphere is said to be
(i) {\it trapped} if $\theta_+\theta_->0$,
(ii) {\it untrapped} if $\theta_+\theta_-<0$
and (iii) {\it marginal} if $\theta_+\theta_-=0$.
(Equivalently, if $g^{-1}(\d r)$ is temporal, spatial or null respectively).
This terminology for surfaces will be extended to hypersurfaces
and space-time regions.
If $\e^f\theta_+\theta_-$ is a function with non-vanishing derivative,
the space-time is divided into trapped and untrapped regions,
separated by marginal hypersurfaces.
The following subdivisions may be made.
\item{(i)} A trapped sphere is {\it future} if $\theta_\pm<0$
and {\it past} if $\theta_\pm>0$ [6--7].
(Equivalently,
if $g^{-1}(\d r)$ is future-temporal or past-temporal respectively).
Future and past trapped spheres occur in black and white holes respectively,
and also in cosmological models.

\item{(ii)} On an untrapped sphere,
a spatial or null normal vector $z$ is {\it outgoing} if $z(\d r)>0$
and {\it ingoing} if $z(\d r)<0$.
Equivalently, fixing the orientation locally by $\theta_+>0$ and $\theta_-<0$,
$z$ is outgoing if $g(z,\p_+)>0$ or $g(z,\p_-)<0$
and ingoing if $g(z,\p_+)<0$ or $g(z,\p_-)>0$ [8].
In particular, $\p_+$ and $\p_-$ are respectively
the outgoing and ingoing null normal vectors.
It is easily checked that
the area is increasing in any outgoing spatial or null direction,
and decreasing in any ingoing spatial or null direction.

\item{(iii)} A marginal sphere with $\theta_+=0$
is {\it future} if $\theta_-<0$,
{\it past} if $\theta_->0$,
{\it bifurcating} if $\theta_-=0$,
{\it outer} if $\p_-\theta_+<0$,
{\it inner} if $\p_-\theta_+>0$
and {\it degenerate} if $\p_-\theta_+=0$ [6--7].
The closure of a hypersurface
foliated by future or past, outer or inner marginal spheres
is called a (non-degenerate) {\it trapping horizon} [6--7].
Outer trapping horizons possess various easily proven properties
which are often intuitively ascribed to black or white holes,
including confinement of observers
and analogues of the zeroth, first and second laws of thermodynamics [6--9].
Inner trapping horizons include cosmological horizons
as well as the possible inner boundaries of black and white holes.
In this article, future (respectively past) outer trapping horizons
will be taken as the definition of black (respectively white) holes.
This enables the discussion of black holes in general space-times,
not just those which are asymptotically flat.
In particular,
there will be no discussion of event horizons or apparent horizons,
which are defined in asymptotically flat space-times only [10--11].
The trapping horizon is a more useful concept
because it is sufficiently general, is defined quasi-locally
and has the properties mentioned above.
\ms\ni
The Misner-Sharp spherically symmetric gravitational energy,
or simply the {\it energy}, may be defined in units $G=1$ by
$$E=\half r\left(1-g^{-1}(\d r,\d r)\right)
=\half r+\e^fr\p_+r\p_-r=\half r+\quart\e^fr^3\theta_+\theta_-.\eqno(4)$$
The form actually given by Misner \& Sharp is derived in Section IV.
Note that $E$ is an invariant.
Indeed, the only invariants of the metric and its first derivatives
are functions of $r$ and $E$, as explained above.
This makes $r$ and $E$ natural variables to use,
as has been rediscovered many times by different authors.
It transpires that remarkably many key geometrical properties
of spherically symmetric space-times are controlled by $r$ and $E$.
\ms\ni
{\it Proposition 1: trapping.}
A metric sphere is trapped if and only if $E>\half r$,
marginal if and only if $E=\half r$,
and untrapped if and only if $E<\half r$.
\ms\ni
{\it Proof.} By definition.
\sq\ms\ni
This property is mathematically trivial given the definition in the above form,
but is physically important because it shows that
the ratio $E/r$ controls the formation of black and white holes,
and trapped spheres generally.
Note that the material (or passive) mass does not have this property;
the sharpest relations [12] between trapped spheres and the material mass
fall short of necessary and sufficient conditions.
In other words, it is not the material mass
which directly controls the formation of black and white holes,
but the effective energy $E$.

Consider the two-dimensional space-time
obtained by taking the quotient by the spheres of symmetry.
If $r=0$ coincides with a boundary of the quotient space-time,
then it will be called a {\it centre}.
A central point $p$ will be called {\it regular} if
$g^{-1}(\d r,\d r)-1=O(r^2)$ as $p$ is approached,
and {\it singular} otherwise.
Then $E=O(r^3)$ at a regular centre.
A singular centre will also be referred to as a {\it central singularity}.

A point in the centre will be said to be {\it trapped}
if it surrounded by a neighbourhood of trapped spheres,
and {\it untrapped} if it surrounded by a neighbourhood of untrapped spheres.
A regular centre is untrapped,
but central singularities may be either trapped,
as in the positive-energy Schwarzschild solution,
or untrapped,
as in the negative-energy Schwarzschild solution.
Whether singularities are trapped or untrapped
is relevant to Penrose's cosmic censorship hypothesis [13--14].
Indeed, one might formulate a local version of weak cosmic censorship [13]
in terms of whether singularities are trapped.
Directly relevant to strong cosmic censorship [14]
is whether the singularity is causal or spatial,
defined with respect to the quotient metric.
It turns out that both features are controlled by $E$ [15--16].
\ms\ni
{\it Proposition 2: central singularities.}
For a central singularity at $p$,
if $E\ge E_0$ in a neighbourhood of $p$ for some constant $E_0>0$
(respectively $E\le E_0\le0$) then:
(i) $p$ is trapped (respectively untrapped);
(ii) if the singularity is differentiable at $p$,
then it is spatial (respectively temporal).
\ms\ni
{\it Proof.}
Consider the case $E\ge E_0>0$ (the case $E\le E_0\le0$ being similar).
(i) In the neighbourhood,
$\half\e^fr^3\theta_+\theta_-=2E-r\ge2E_0-r>0$ for sufficiently small $r<2E_0$.
So $\theta_+\theta_->0$ in a neighbourhood of $p$,
i.e.\ this neighbourhood consists of trapped spheres.
(ii) The tangent vector $z=\p/\p\zeta$ to the singularity
is a linear combination $z=\beta\p_+-\alpha\p_-$ of the null normals $\p_\pm$,
so that $0=\p r/\p\zeta=\beta\p_+r-\alpha\p_-r$.
But $\p_+r\p_-r=\quart r^2\theta_+\theta_->0$ in the above neighbourhood,
so $\alpha\beta>0$, which means that $z$ is spatial.
\sq\ms\ni
This reflects a physically important idea:
there is a connection between the sign of the energy,
the causal nature of singularities and whether they are trapped.
Combined with the property $E\ge0$,
which will be derived under certain assumptions in Proposition 6,
this result supports the cosmic censorship hypothesis.
It falls short of a proof of cosmic censorship for two reasons.
Firstly, it is possible to have $E<0$
when the assumptions of Proposition 6 do not hold,
as will be discussed at the end of Section III.
Secondly, there is the case where $E|_p=0$,
in which case $p$ could be a spatial, null or temporal singularity,
or a regular centre.
Specifically, if $2E/r|_p>1$ the singularity is spatial and trapped,
and if $2E/r|_p<1$ the singularity is temporal and untrapped.
If $2E/r|_p=1$, one must look at higher orders, $(2E/r-1)/r$, and so on.
Exactly which possibility occurs seems to depend on the matter field.
According to the analysis of Christodoulou [17], for a massless scalar field
it is possible to obtain causal central singularities,
but such configurations are non-generic with respect to initial data.
Conversely, for pure radiation (or null dust),
a sufficiently weak wave travelling into an initially flat space-time
necessarily creates a null singularity
which is at least locally visible [18--19].
Such visible null singularities are also possible for dust [20--22].

Despite such material-dependent differences, one useful fact remains:
if $E\ge0$, a central singularity which is either causal or untrapped
must have vanishing energy, $E|_p=0$.
(For instance,
the analysis of Joshi \& Dwivedi [23] concerns such singularities).
This at least constrains counter-examples to cosmic censorship.
Moreover, one might expect that
zero-energy singularities are unstable or non-generic in some sense.
Similarly, it has been suggested [24] that
zero-energy singularities are non-gravitational
and do not conflict with the spirit of the cosmic censorship hypothesis.
Beyond spherical symmetry,
there is also some evidence for a weakened form of cosmic censorship
in which positive-energy singularities are censored [25].
\ms\ni
{\it Proposition 3: asymptotics.}
In an asymptotically flat space-time,
$E$ coincides with the Bondi-Sachs (scalar) energy $E_{BS}$ at null infinity,
and with the Arnowitt-Deser-Misner (scalar) energy $E_{ADM}$
at spatial infinity.
\ms\ni
{\it Proof.}
By definition [2--3,11,15,26,27]:
$$E_{BS}=\lim_{A\to\infty}{1\over{8\pi}}\sqrt{{A\over{16\pi}}}\int\mu
(\R+\e^f\theta_+\theta_-)\eqno(5)$$
where $\R$ is the Ricci scalar, $\mu$ the area form and $A=\int\mu$ the area
of a family of affinely parametrised surfaces
lying in a null hypersurface approaching $\scri^\pm$.
In spherical symmetry, $\int\mu=4\pi r^2$ and $\R=2/r^2$, so that
$$E_{BS}=\l E\eqno(6)$$
for a null curve $\gamma$ approaching $p\in\scri^\pm$.
Similarly, by definition [1,15,28]:
$$E_{ADM}=\lim_{A\to\infty}{1\over{8\pi}}\sqrt{{A\over{16\pi}}}\int\mu
(\R+\e^f\theta_+\theta_--\half\e^f\langle\sigma_+,\sigma_-\rangle)\eqno(7)$$
for a family of surfaces parametrised by area radius $r=\sqrt{A/4\pi}$
lying in a spatial hypersurface approaching $i^0$,
where $\sigma_\pm$ are the shears corresponding to the expansions $\theta_\pm$.
In spherical symmetry, $\sigma_\pm=0$, so that
$$E_{ADM}=\l E\eqno(8)$$
for a spatial curve $\gamma$ approaching $p=i^0$.
\sq\ms\ni
The result shows that the asymptotic energies (in spherical symmetry)
are just special cases of $E$,
defined at infinity in an asymptotically flat space-time.
It is usual to interpret the asymptotic energies
as measuring the total energy (or mass---see Appendix C)
of an asymptotically flat space-time,
whereas $E$ provides a more general definition of energy
which applies locally as well as asymptotically.
\bs\ni
{\bf III. Dynamical properties}
\ms\ni
Having derived various purely geometrical properties of $E$,
consider now applying the Einstein equations.
The most general form of the Einstein tensor in spherical symmetry
determines the most general material stress-energy tensor $T$,
given by the line-element
$$pr^2\d\Omega^2+T_{++}(\d\xi^+)^2+T_{--}(\d\xi^-)^2+2T_{+-}\d\xi^+\d\xi^-.
\eqno(9)$$
The Einstein equations are
$$\eqalignno
{&\p_\pm\p_\pm r+\p_\pm f\p_\pm r=-4\pi rT_{\pm\pm}&(10a)\cr
&r\p_+\p_-r+\p_+r\p_-r+\half\e^{-f}=4\pi r^2T_{+-}&(10b)\cr
&r^2\p_+\p_-f+2\p_+r\p_-r+\e^{-f}=8\pi r^2(T_{+-}+\e^{-f}p)&(10c)\cr}$$
recalling the units $G=1$.
The variation of $E$ is determined by these equations as
$$\p_\pm E=4\pi\e^fr^2(T_{+-}\p_\pm r-T_{\pm\pm}\p_\mp r)
=2\pi\e^fr^3(T_{+-}\theta_\pm-T_{\pm\pm}\theta_\mp).\eqno(11)$$
This can also be written in a manifestly covariant form [29--30].
\ms\ni
{\it Proposition 4: vacuum.}
In vacuo, $E$ is constant
and the solution is locally isometric to the Schwarzschild solution
with energy $E$.
\ms\ni
{\it Proof.}
In vacuo, $\p_\pm E=0$, so $E$ is constant.
The vacuum Einstein equations are
$$\eqalignno
{&\p_\pm(\e^f\p_\pm r)=0&(12a)\cr
&\p_+\p_-r=-E\e^{-f}r^{-2}&(12b)\cr
&\p_+\p_-f=-2E\e^{-f}r^{-3}.&(12c)\cr}$$
If $E\not=0$,
a straightforward calculation yields $\p_+\p_-(f-\log r-r/2E)=0$,
with general solution $f=\log r+r/2E+f^++f^-$
for some functions $f^\pm(\xi^\pm)$ of integration.
The rescaling freedom (2) corresponds to choice of $f^\pm$,
so that one may fix coordinates such that
$$\e^f=r\e^{r/2E}/16E^3.\eqno(13)$$
The equations (12a) integrate to $\e^f\p_\pm r=\eta^\mp$
for some functions $\eta^\pm(\xi^\pm)$ of integration.
Differentiating using (12b), $\p_\pm\eta^\pm=-1/4E$,
which integrates to $\eta^\pm=-\xi^\pm/4E$, fixing the zero. Thus
$$\xi^+\xi^-=(1-r/2E)\e^{r/2E}\eqno(14)$$
which implicitly determines $r$ (and hence $f$) as a function of $\xi^+\xi^-$.
This is the Kruskal form of the Schwarzschild solution with energy $E$,
which can be put in static form in terms of $t^\pm=2E\log(-\xi^\pm/\xi^\mp)$.
Similarly, if $E=0$ then flat space-time is recovered.
\sq\ms\ni
This is a proof of Birkhoff's theorem:
a vacuum, spherically symmetric space-time must be the Schwarzschild solution.
The proof improves on the usual one [10]
in that a global coordinate chart is obtained automatically,
so that one does not have to subsequently join
the $r>2E$ and $r<2E$ Schwarzschild charts.

The Schwarzschild solution provides an example of Propositions 1--3.
For the case $E>0$ there are trapped spatial singularities,
while for the case $E<0$ there is an untrapped temporal singularity.
In the former case, there are trapped spheres
in the black-hole and white-hole regions $E>\half r$,
and the event horizons coincide with the trapping horizons $E=\half r$.

Consider now the non-vacuum cases.
In order to obtain results which are as general as possible,
the type of matter will not be fixed
but energy conditions will be imposed instead.
Three useful energy conditions are as follows [10--11].
The null energy (or convergence) condition states that
a ``null observer'' measures non-negative energy:
$$\hbox{NEC:}\qquad g(u,u)=0\quad\Rightarrow\quad T(u,u)\ge0.\eqno(15)$$
The weak energy condition states that
a causal observer measures non-negative energy:
$$\hbox{WEC:}\qquad g(u,u)\le0\quad\Rightarrow\quad T(u,u)\ge0.\eqno(16)$$
The dominant energy condition states that
a future-causal observer measures future-causal momentum:
$$\hbox{DEC:}\qquad g(u,u)\le0,g(v,v)\le0,g(u,v)\le0
\quad\Rightarrow\quad T(u,v)\ge0.\eqno(17)$$
Clearly
$$\hbox{DEC}\quad\Rightarrow\quad\hbox{WEC}\quad\Rightarrow\quad\hbox{NEC}.
\eqno(18)$$
All this applies to general space-times.
In the spherically symmetric case,
$$\hbox{NEC}\quad\Rightarrow\quad T_{\pm\pm}\ge0\eqno(19)$$
and
$$\hbox{DEC}\quad\Rightarrow\quad T_{\pm\pm}\ge0,T_{+-}\ge0.\eqno(20)$$
\ms\ni
{\it Proposition 5: monotonicity.}
If the dominant energy condition holds on an untrapped sphere,
$E$ is non-decreasing (respectively non-increasing)
in any outgoing (respectively ingoing) spatial or null direction.
\ms\ni
{\it Proof.}
Fix the orientation of the untrapped sphere by $\theta_+>0$ and $\theta_-<0$.
The variation formula (11) and dominant energy condition (20) yield
$\p_+E\ge0$ and $\p_-E\le0$,
i.e.\ $E$ is non-decreasing (respectively non-increasing)
in the outgoing (respectively ingoing) null direction.
If $z=\p/\p\zeta$ is an outgoing spatial vector,
then $z=\beta\p_+-\alpha\p_-$ with $\alpha>0$ and $\beta>0$,
which yields $\p E/\p\zeta=\beta\p_+E-\alpha\p_-E\ge0$.
Similarly for an ingoing spatial direction.
\sq\ms\ni
The proof illustrates the economy of the double-null approach:
monotonicity in any spatial direction
follows immediately from monotonicity in the null directions.
This monotonicity property has the physical interpretation that
the energy contained in a sphere is non-decreasing
as the sphere is perturbed outwards.
Note that the result is for untrapped spheres only,
though a similar result for marginal spheres will be given in Proposition 8.
There is no possibility of a similarly general monotonicity result
for trapped spheres, since they do not have a locally preferred orientation.
\ms\ni
{\it Proposition 5A: asymptotic monotonicity.}
If the dominant energy condition holds at $\scri^+$ (respectively $\scri^-$),
then $E_{BS}$ is non-increasing (respectively non-decreasing) to the future.
\ms\ni
{\it Proof.}
By Propositions 3 and 5, $\p_-E_{BS}=\p_-(\l E)=\l\p_-E\le0$.
Here commutativity follows from the asymptotic expansions [11], as in [8].
\sq\ms\ni
At $\scri^+$, this is the Bondi-Sachs energy-loss property,
which is usually interpreted as
describing a loss of energy due to outgoing radiation.
Similarly, the more general monotonicity property of $E$ may be interpreted as
being due to ingoing and outgoing radiation.

Positivity of $E$ is already known in some circumstances [30]
but can be established directly and quite generally
from the monotonicity property.
Note first that $E$ may be negative,
as in the negative-energy Schwarzschild solution.
On the other hand, Proposition 1 shows that
$E$ is automatically positive for trapped and marginal spheres,
so that it is only for untrapped spheres that $E$ might be negative.
Positivity on untrapped spheres is established below
in two physically relevant contexts,
namely where there is a regular centre
and where there is a black or white hole.
\ms\ni
{\it Proposition 6: positivity.}
If the dominant energy condition holds
on an untrapped spatial hypersurface with regular centre,
then $E\ge0$ on the hypersurface.
\ms\ni
{\it Proof.}
By Proposition 5, since $E=0$ at a regular centre.
\sq\ms\ni
Moreover, denoting the tangent vector to the hypersurface as before by
$\p/\p\zeta=\beta\p_+-\alpha\p_-$, with $\zeta=0$ being the centre,
one can write explicitly
$$E(\zeta)=4\pi\int_0^\zeta\e^fr^2
\{(\beta T_{+-}+\alpha T_{--})\p_+r-(\alpha T_{+-}+\beta T_{++})\p_-r\}
\d\hat\zeta.\eqno(21)$$
The positivity property has the physical interpretation that
under the stated circumstances, total energy cannot be negative.
This is not immediately obvious even given an energy condition on the matter,
since gravitational potential energy tends to be negative.
The result shows that the total energy $E$, including potential energy,
cannot be negative.
\ms\ni
{\it Proposition 7: area inequality.}
If the dominant energy condition holds on an untrapped spatial hypersurface
bounded at the inward end by a marginal sphere of radius $r_0$,
then $E\ge\half r_0$ on the hypersurface.
\ms\ni
{\it Proof.}
By Propositions 1 and 5.
\sq\ms\ni
Since $r_0>0$, this is a stronger result than mere positivity of $E$:
there is a positive lower bound on $E$.
The physical interpretation is that
if there is a black or white hole of area $4\pi r_0^2$,
then the energy measured outside the hole is at least $\half r_0$.
As for the positivity result, one can write an explicit formula
$$E(\zeta)=\half r_0+4\pi\int_{\zeta_0}^\zeta\e^fr^2
\{(\beta T_{+-}+\alpha T_{--})\p_+r-(\alpha T_{+-}+\beta T_{++})\p_-r\}
\d\hat\zeta\eqno(22)$$
where $\zeta=\zeta_0$ is the marginal surface.
\ms\ni
{\it Proposition 6A: asymptotic positivity.}
If the dominant energy condition holds
on a spatial hypersurface which has a regular centre
and extends to $\scri^\pm$ (respectively $i^0$),
then $E_{BS}\ge0$ (respectively $E_{ADM}\ge0$) there.
\ms\ni
{\it Proof.}
If the hypersurface is untrapped, Propositions 3 and 6 suffice.
Otherwise, the hypersurface contains an outermost marginal sphere,
so Propositions 3 and 7 suffice.
\sq\ms\ni
This combines the famous positive-energy theorems
for the spherically symmetric case.
Note that the energy can be negative if the above assumptions do not hold.
For instance, the asymptotic energies are negative
for the negative-energy Schwarzschild solution.
\ms\ni
{\it Proposition 7A: asymptotic area inequality.}
If the dominant energy condition holds on a spatial hypersurface
which contains an outermost marginal sphere of radius $r_0$
and which extends to $\scri^\pm$ (respectively $i^0$),
then $E_{BS}\ge\half r_0$ (respectively $E_{ADM}\ge\half r_0$) there.
\ms\ni
{\it Proof.}
By Propositions 3 and 7.
\sq\ms\ni
The result is the spherically symmetric case of the isoperimetric inequality
conjectured by Penrose [5],
who argued that it was required by the cosmic censorship hypothesis.
Establishing the inequality even in spherical symmetry appears to be new.
It was recently established
for maximal hypersurfaces in spherical symmetry [31].

The properties 5A, 6A and 7A of the asymptotic energies
are of interest in their own right.
Nevertheless, they are just special cases of properties of $E$.
If these properties of the asymptotic energies
are accorded their usual conceptual and physical importance,
then the more general properties of $E$ are of even greater importance.
The idealisation of asymptotic flatness
is no longer necessary for the formulation of such ideas about energy.
\ms\ni
{\it Proposition 8: second law} [6].
If the null energy condition holds
on a future (respectively past) outer trapping horizon,
or on a past (respectively future) inner trapping horizon,
then $E=\half r$ is non-decreasing (respectively non-increasing)
along the horizon.
\ms\ni
{\it Proof.}
Denote the tangent to the horizon by $\p/\p\zeta=\beta\p_+-\alpha\p_-$
and fix the orientations by $\theta_+=0$ and $\beta>0$ on the horizon.
Then $0=\p\theta_+/\p\zeta=\beta\p_+\theta_+-\alpha\p_-\theta_+$ yields
$\p r/\p\zeta=-\alpha\p_-r=-\beta r\theta_-\p_+\theta_+/2\p_-\theta_+$.
The focussing equation (10a) and null energy condition (19) yield
$\p_+\theta_+\le0$,
and the signs of $\theta_-$ and $\p_-\theta_+$ are given by
the definition of future or past, outer or inner trapping horizons.
Thus $\p r/\p\zeta\ge0$ for future outer or past inner trapping horizons,
and $\p r/\p\zeta\le0$ for past outer or future inner trapping horizons.
\sq\ms\ni
Propositions 5--8 may be loosely summarised as follows.
If the dominant energy condition holds,
$E$ is non-decreasing in outgoing directions,
defined for untrapped or marginal surfaces,
including at conformal infinity if it exists.
So in the untrapped region outside a regular centre, $E$ is non-negative,
including at conformal infinity if it exists for the untrapped region.
But $E$ is also positive in a trapped region.
Moreover, in the untrapped region outside a black or white hole
(defined by outer marginal surfaces) of area $4\pi r_0^2$,
$E\ge\half r_0$,
including at conformal infinity if it exists for the untrapped region.

Propositions 1--8 alone
give a quite coherent picture of gravitational collapse,
which may be further refined using related results [6--9].
Recall first the expected picture
according to the cosmic censorship hypothesis [13--14].
Here a centre which is initially regular subsequently becomes singular,
but in such a way that the singularity is generically spatial and trapped.
That is, a future trapping horizon forms before the singularity.
More precisely, one expects the trapping horizon to intersect the centre,
typically at the first singular point.\footnote\dag
{This does not contradict the fact that,
when using a family of spatial hypersurfaces developing in time,
trapped surfaces are often first detected away from the centre.
This just depends on the choice of hypersurfaces; if the horizon is spatial,
one may choose a spatial hypersurface to touch it anywhere.
There is no unique first moment of trapped surface formation,
except in a limiting sense at the centre.}
The horizon develops outwards with non-decreasing $E=\half r$.
Outside the horizon is an untrapped region in which $0\le E<\half r$.
Inside the horizon is a trapped region in which $E>\half r$,
which is expected to extend to the centre $r=0$
in such a way that $E$ is generically positive there.
This picture has been confirmed for the massless scalar field
by Christodoulou [17].

To what extent do the above results support this picture?
Firstly, there is the connection of Proposition 2 between
the causal nature of central singularities and whether they are trapped.
Thus two aspects of cosmic censorship are linked.
Secondly, both aspects are linked to the sign of the energy $E$,
again a physically important connection.
Thirdly, there are the positivity properties of $E$,
Propositions 1, 6 and 7.
Consider then a space-time with partially regular centre,
and the external region $X$ of points
connectable to the regular part of the centre
by a non-temporal curve (respecting the spherical symmetry).
As long as $X$ remains untrapped,
$E$ is positive and no central singularities can form.
If part of $X$ is trapped, i.e.\ a trapping horizon forms,
then inside the trapped region
any central singularity must be either spatial or have zero energy.
One expects that
zero-energy singularities are non-generic or unstable in some sense.
The only other way that a non-spatial central singularity can form in $X$
is if a second trapping horizon forms,
separating the trapped region from another untrapped region.
That is, the black or white hole must have an inner boundary.
This is also possible, as the Reissner-Nordstr\"om solution shows,
but again such horizons are thought to be unstable
by the energy-inflation effect [32--34]:
for a test field on such a background,
the perturbation in $E$ typically becomes unbounded at the horizon.
This leaves the only other possibility for non-spatial central singularities
as being outside $X$.
Again this is possible:
the centre may simply become singular but remain causal,
implying that $E$ is no longer positive.
There are such examples [17--24],
but it is noteworthy that they have zero rather than negative $E$.
This suggests that there may be some mechanism forbidding
such formation of negative-energy singularities.
As to the zero-energy singularities,
one might again expect them to be unstable in some sense.
Both possibilities indicate the need to study the behaviour of $E$
as the centre becomes singular.
Finally, there is the possibility of non-central singularities,
such as shell-crossing singularities [35--36].
However, these have been found only for matter fields
which admit similar singularities as test fields on flat space-time,
and therefore may be dismissed as pathological matter models.

To conclude:  to find a convincing counter-example to cosmic censorship,
i.e.\ the stable formation of a causal singularity,
one needs to study the possibility of negative-energy singularities,
the stability of zero-energy singularities
or the stability of inner horizons.
It is noteworthy that $E$ plays a crucial role in each case.
The importance of energy for cosmic censorship seems clear.
\bs\ni
{\bf IV. Spatial hypersurfaces and the Newtonian limit}
\ms\ni
Consider any spatial hypersurface $\Sigma$.
Set up coordinates $(\tau,\zeta)$ such that
$\p/\p\zeta$ is tangent to $\Sigma$
and $\p/\p\tau$ is orthogonal to $\p/\p\zeta$,
with $\tau$ being proper time:
$$\eqalignno
{&g(\p/\p\tau,\p/\p\zeta)=0&(23a)\cr
&g(\p/\p\tau,\p/\p\tau)=-1.&(23b)\cr}$$
Define a function $\lambda$ by
$$\e^\lambda=g(\p/\p\zeta,\p/\p\zeta).\eqno(24)$$
For any such coordinates $(\tau,\zeta)$,
the rescaling freedom (2) can be used to fix $\xi^\pm$ such that
$$\sqrt2\e^{-f/2}\d\xi^\pm=\d\tau\pm\e^{\lambda/2}\d\zeta.\eqno(25)$$
The line-element (1) transforms to
$$\d s^2=r^2\d\Omega^2+\e^\lambda\d\zeta^2-\d\tau^2\eqno(26)$$
where $r$ and $\lambda$ are functions of $(\tau,\zeta)$.
Denote $\dot\varphi=\p\varphi/\p\tau$
and $\varphi'=\p\varphi/\p\zeta$.
Then the definition (4) of $E$ can be rewritten as
$$1-{2E\over{r}}=\e^{-\lambda}(r')^2-\dot r^2\eqno(27)$$
which is the form actually given by Misner \& Sharp [4],
with different notation.

The Einstein equations may be transformed to these coordinates,
but for the following results
it suffices to find the corresponding variation formulas for $E$, which are
$$\eqalignno
{&E'=4\pi r^2(T_{00}r'-T_{01}\dot r)&(28a)\cr
&\dot E=4\pi r^2\e^{-\lambda}(T_{01}r'-T_{11}\dot r)&(28b)\cr}$$
where
$$\eqalignno
{&T_{00}=T(\p/\p\tau,\p/\p\tau)&(29a)\cr
&T_{01}=T(\p/\p\tau,\p/\p\zeta)&(29b)\cr
&T_{11}=T(\p/\p\zeta,\p/\p\zeta).&(29c)\cr}$$
\ms\ni
{\it Proposition 9: small spheres.}
Near a regular centre with tangent $\p/\p\tau$,
if $T_{00}=O(1)$ and $T_{01}=O(1)$ then
$$E=\ft\pi r^3T_{00}|_{r=0}+O(r^4).\eqno(30)$$
\ms\ni
{\it Proof.}
Near a regular centre $r=0$, $r'=O(1)$.
If $\p/\p\tau$ is tangent to the centre, then $\dot r=O(r)$.
Thus $E'=4\pi r^2T_{00}r'+O(r^3)$,
which integrates along the hypersurface $\Sigma$ to the above result.
\sq\ms\ni
In other words, the leading term in $E$
is the product of volume $\ft\pi r^3$ and density $T_{00}$,
as would be expected physically.

Misner \& Sharp [4] derived a useful formula for $E$
which will be derived below and used to find the Newtonian limit.
Misner \& Sharp considered a perfect fluid
with energy density $\rho$, pressure $p$ and velocity $\p/\p\hat\tau$:
$$\eqalignno
{&T=(\rho+p)\d\hat\tau\otimes\d\hat\tau+pg&(31a)\cr
&g(\p/\p\hat\tau,\p/\p\hat\tau)=-1.&(31b)\cr}$$
Adapt the $(\tau,\zeta)$ coordinates to the fluid by taking $\tau=\hat\tau$.
Then the variation formulas (28) for $E$ reduce to
$$\eqalignno
{&E'=4\pi r^2\rho r'&(32a)\cr
&\dot E=-4\pi r^2p\dot r.&(32b)\cr}$$
The second equation (32b) expresses the rate $\dot E$ of work done
by the force $4\pi r^2p$ due to pressure.
In the absence of pressure, $E$ is conserved, $\dot E=0$.
The first equation (32a)
allows $E$ to be expressed as an integral of $\varepsilon$ over $\Sigma$.
Specifically, if $\Sigma$ has a regular centre $\zeta=0$, (32a) integrates to
$$E(\zeta)=\int_0^\zeta4\pi r^2\rho r'\d\hat\zeta.\eqno(33)$$
Since the volume form $*1$ of $\Sigma$ is given by (26) as
$$\I\varphi=\int_0^\zeta4\pi r^2\e^{\lambda/2}\varphi\,\d\hat\zeta\eqno(34)$$
this may be rewritten as
$$E=\I\rho\e^{-\lambda/2}r'
=\I\rho\left(1+\dot r^2-{2E\over{r}}\right)^{1/2}\eqno(35)$$
where the second expression follows from (27).
Comparing with the material (or passive) mass
$$M=\I T_{00}=\I\rho\eqno(36)$$
it can be seen that the integrand for $E$ differs from that of $M$ by a factor
which Misner \& Sharp interpreted as being due to kinetic and potential energy.
This can be made precise in the Newtonian limit in terms of
the kinetic energy $K$ and gravitational potential energy $V$,
defined as explained in Appendix A by
$$\eqalignno
{&K=\I\half\rho\dot r^2&(37a)\cr
&V=-\I{M\rho\over{r}}.&(37b)\cr}$$
Factors of the speed of light $c$ may be introduced on dimensional grounds
by the formal replacements $\tau\mapsto c\tau$,
$(r,*1)\mapsto(r,*1)$,
$(\rho,M)\mapsto c^{-2}(\rho,M)$,
$(p,K,V,E)\mapsto c^{-4}(p,K,V,E)$, assumed in the remainder of this Section.
These factors are determined simply by
the desired interpretation of the various quantities as time, length, mass etc.
\ms\ni
{\it Proposition 10: Newtonian limit.}
For a perfect fluid on a spatial hypersurface $\Sigma$ with regular centre:
if $(*1,r,\dot r,\rho)=O(1)$ as $c\to\infty$,
then $(M,K,V)=O(1)$ and
$$E=Mc^2+K+V+O(c^{-2}).\eqno(38)$$
Additionally, if $p=O(1)$ then Newtonian conservation of mass is recovered:
$$\dot M=O(c^{-2}).\eqno(39)$$
\ms\ni
{\it Proof.}
Inserting the factors of $c$, (36) and (37) take the same form,
so that $(M,K,V)=O(1)$,
while (35) takes the form
$$E=c^2\I\rho
\left(1+{\dot r^2\over{c^2}}-{2E\over{c^4r}}\right)^{1/2}.\eqno(40)$$
Thus $E=c^2\I\rho+O(1)=Mc^2+O(1)$ to leading order.
Expanding the square root,
$E=Mc^2+\I\rho(\half\dot r^2-E/rc^2)+O(c^{-2})=Mc^2+K+V+O(c^{-2})$.
Similarly, (32b) expands to
$-4\pi r^2p\dot r=\dot Mc^2+O(1)$.
\sq\ms\ni
In words, $E$ yields the Newtonian mass $M$ to leading order
and the Newtonian kinetic energy $K$ and gravitational potential energy $V$
to next highest order.
This illustrates how $E$ measures the combined energy
including contributions from mass, kinetic energy and potential energy.
Note also that the quantities $M$, $K$ and $V$
are all defined in the full theory rather than just the Newtonian limit.
Of these, $M$ may always be interpreted as the material (or passive) mass,
whereas the interpretation of $K$ and $V$ as kinetic and potential energy
is reliable in the Newtonian limit only.
In general,
$E$ cannot be expressed as a sum of individually meaningful energies,
as the form (35) indicates.
It is only the combined energy $E$ which is physically meaningful in general.

Incidentally, the above considerations
provide one reason for referring to $E$ as an energy rather than a mass,
though the latter is more common.
Although mass and energy are formally equivalent in Relativity,
the two words carry connotations
inherited from their status as distinct concepts in Newtonian theory.
Specifically, mass is a measure of matter
whereas energy exists in many different forms.
So it is reasonable to describe $M$ as mass,
since it is simply the integral (36) of the material density,
and to describe $E$ as energy,
since it contains contributions from kinetic and potential energy.

Under the assumptions of Proposition 10,
the inverse metric is Euclidean to leading order,
$g^{-1}=\delta^{-1}+O(c^{-2})$,
where the flat metric $\delta$ is given by $r^2\d\Omega^2+\d r^2$,
so that $r$ is a standard radial coordinate.
Thus the flat space of Newtonian theory is recovered.

Note that it has not been necessary
to introduce the usual coordinate conditions [37]
required to obtain an inertial frame in which Newton's laws hold,
or to obtain Newtonian solutions from solutions to the Einstein equations [38].
In contrast, the Newtonian behaviour of $E$ may be obtained simply by
expanding a formula for $E$ in powers of $c$.
In this sense,
the recovery of Newtonian mass and energy from $E$ is more robust than
the complete recovery of Newton's gravitational theory from Einstein's theory.
This could be interpreted as meaning that energy is a fundamental concept
which connects Newtonian theory with Relativity.
Certainly this is consistent with the key role
that the equivalence of mass and energy played
in the historical development of Relativity.
\bs\ni
{\bf V. Test particles and the special-relativistic limit}
\ms\ni
The energy of a spherical shell of test particles may be defined as
the variation in $E$ induced by the perturbation of the test particles,
as suggested by H\'aj\'\i\v cek [39].
Denote the perturbations of $(T,E,M)$ by $(\T,\E,\M)$.
In terms of the constant rest mass $m$ and velocity $\p/\p\hat\tau$
of the shell,
the stress-energy tensor is given analogously to (31) by
$$\T=m\delta\,\d\hat\tau\otimes\d\hat\tau\eqno(41)$$
where $\delta$ is the Dirac distribution with support at
the intersection $S$ of the shell with a spatial hypersurface $\Sigma$:
$$\I\delta\varphi=\varphi|_S.\eqno(42)$$
Evaluation on $S$ is clear in context and will be omitted henceforth.
Take $\Sigma$ to be orthogonal to the shell
and adopt the notation of Section IV for spatial hypersurfaces,
identifying $\tau=\hat\tau$.
Then the H\'aj\'\i\v cek energy $\E$ is defined by
the perturbation of (11) or (28),
$$\eqalignno
{&\E'=4\pi r^2(\T_{00}r'-\T_{01}\dot r)&(43a)\cr
&\dot\E=4\pi r^2\e^{-\lambda}(\T_{01}r'-\T_{11}\dot r).&(43b)\cr}$$
and $\M$ by (36),
$$\M=\I\T_{00}.\eqno(44)$$
\ms\ni
{\it Proposition 11: test particles.}
For a spherical shell of test particles,
(i) the perturbations in $M$ and $E$ are respectively
$$\eqalignno
{&\M=m&(45a)\cr
&\E=\pm m\left(1+\dot r^2-{2E\over{r}}\right)^{1/2}&(45b)\cr}$$
where the sign is that of $r'$.
Both are conserved:
$$\eqalignno
{&\dot\M=0&(46a)\cr
&\dot\E=0.&(46b)\cr}$$
(ii) In the Newtonian limit, with assumptions as for Proposition 10,
$$\pm\E=mc^2+\half m\dot r^2-{mM\over{r}}+O(c^{-2}).\eqno(47)$$
\ms\ni
{\it Proof.}
(i) Since $\T_{00}=m\delta$ and $\T_{01}=\T_{11}=0$,
(44) reads $\M=\I m\delta=m$,
(43b) reads $\dot\E=0$
and (43a) reads $\E'=4\pi r^2r'm\delta$,
which integrates using (27) and (34) to
$\E=\pm\I m\delta(1+\dot r^2-2E/r)^{1/2}=\pm m(1+\dot r^2-2E/r)^{1/2}$.
(ii) Inserting the factors of $c$,
$$\pm\E=mc^2
\left(1+{\dot r^2\over{c^2}}-{2E\over{c^4r}}\right)^{1/2}\eqno(48)$$
which expands as above.
\sq\ms\ni
In words:  the perturbation in $M$ is just the rest mass $m$,
while the perturbation in $E$ contains contributions from energy,
in particular the correct Newtonian expressions for
kinetic energy $\half m\dot r^2$ and gravitational potential energy $-mM/r$.
Combined with the fact that $\E$ is conserved as the shell evolves,
this confirms the physical interpretation of $\E$
as the active energy of the shell of test particles.

Conservation (46b) of $\E$, written explicitly using (45b),
is equivalent to the equation of motion of the test particles:
$$\ddot r=-\dot\Phi/\dot r\eqno(49)$$
where $\Phi=-E/r$. This has a similar form to the Newtonian equation of motion
of test particles in a central gravitational field, with $E$ playing the role
of the active gravitational energy of the source and $\Phi$ the gravitational
potential. Again this confirms the physical interpretation of $E$ as the active
gravitational energy.
\ms\ni
{\it Proposition 12: special-relativistic limit.}
In flat space-time, the perturbation in $E$ for a spherical shell of test
particles is
$$\E|_{E=0}=\pm m\left(1+\dot r^2\right)^{1/2}.\eqno(50)$$
\ni
{\it Proof.}
By Proposition 11.
\sq\ms\ni
This has been dignified as a separate Proposition because it is important that
$E$ yields the correct result in the special-relativistic limit,
namely that $\E$ reduces to the standard expression (50)
for the energy of test particles in special relativity.
In particular, $\E$ is not just the rest mass.
This is noteworthy since the opposite is sometimes claimed,
namely that $E$ corresponds to the rest mass rather than energy
of a test particle in flat space-time.
This misunderstanding seems to be the reason---other than apathy---that
$E$ is usually described as mass rather than energy.
On the contrary,
using the terminology standard in both special relativity and Newtonian theory,
$\E$ and $E$ are energies and $m$ and $M$ masses.
In the same sense, the following are all energies rather than masses:
quasi-local energies which agree with $E$ in spherical symmetry,
such as those of Penrose and Hawking [15,40--46];
the asymptotic (scalar) energies $E_{ADM}$ and $E_{BS}$;
the Schwarzschild energy.

The mistaken claim that $E$ corresponds to $m$ rather than $\E$
is sometimes ascribed to the fact that, in flat space-time,
$\E$ depends on the choice of frame (or observer)
whereas $m$ is frame-independent, like $E$.
But except in flat space-time, $\E$ is also frame-independent:
the velocity $\dot r$ is relative to the preferred frame determined by $r$,
not the frame of some observer.
Only in flat space-time is $r$ not unique, due to the Lorentz symmetry.
It is this degeneracy of flat space-time that renders $\E$ frame-dependent
in that special case.
Frame-dependence is not a general feature of energy,
as discussed further in Appendix C.
For instance, Newtonian gravitational potential energy is frame-independent.
\bs\ni
{\bf VI. Remarks: gravitational energy in general}
\ms\ni
The Misner-Sharp energy $E$ has an impressive variety of useful properties,
ranging from the asymptotically flat, vacuum, small-sphere,
Newtonian, test-particle and special-relativistic limits
to the black holes and singularities
characteristic of strong gravitational fields.
These properties have an exact geometrical character
and are simultaneously of direct physical relevance.
In particular, $E$ has quite general monotonicity and positivity properties,
determines the causal nature of central singularities,
and determines when trapping occurs.
This makes $E$ useful in many different applications,
particularly regarding black holes and singularities.
Indeed, $E$ has been rediscovered by many authors
and is often crucial to their analyses [16--22,24,29--30,32--34,36,39].
In particular, $E$ arises as the charge
associated with a conserved current, as explained in Appendix B,
and can be derived from the ``2+2'' Hamiltonian [15].
One might also speculate that the radius $r$ and energy $E$
are relevant in quantum theory [47].
Certainly $E$ plays a key role in the vacuum case,
according to the analysis of Kucha\v r [48].

The conceptual and physical importance of $E$ in spherical symmetry
encourages the search for a more general definition of gravitational energy.
It is widely accepted that such a general definition
should reduce to $E$ in spherical symmetry,
as occurs for definitions in the style of Penrose [40--44]
or Hawking [15,45--46].
Exceptions include the Brown-York energy [49],
which gives a value different from $E$ for the Schwarzschild solution,
and the Bartnik energy [50], which is undefined for trapped surfaces.
Whether such definitions have some other physical meaning is unclear,
but they do not represent gravitational energy
in the sense of Propositions 1--12.

Familiarity with the spherically symmetric case
also yields other guidelines to more general definitions.
Firstly,
it is noteworthy that $E$ is defined on spheres rather than hypersurfaces.
One can write an implicit expression for $E$
involving an integral (35) over a hypersurface with regular centre,
if such a hypersurface exists,
but the general definition (4) or (27) is a function of spheres only.
This is quite different to the definition of mass $M$
as an integral (36) over a hypersurface.
So rather than looking for a definition of gravitational energy
as a hypersurface integral, as would be natural in Newtonian theory,
one should look for a surface integral.
Specifically, one wants an invariant of the intrinsic and extrinsic curvature
of an embedded surface,
which following Penrose [40] may be called quasi-local energy.

Another guideline to more general definitions
is that one would like analogous positivity and monotonicity properties,
and analogous relations to black holes
and possibly to singularities which are central in some sense.
The Hawking energy [45]
takes the value $\sqrt{A/16\pi}$ on a spherical marginal surface of area $A$,
and so automatically generalises Propositions 1 and 8.
The Hawking energy also has the same small-sphere behaviour
as in Proposition 9 [51].
Generalisations of the positivity and monotonicity Propositions 5--7
for the Hawking energy have also been found [8].
A warning should be sounded that such positivity and monotonicity theorems
involve certain assumptions without which the conclusions are invalid.
Even the Schwarzschild energy may be negative.
Thus there is little point in
searching for a definition of gravitational energy
which is always non-negative by definition,
as is sometimes suggested [50].

The final lesson of the spherically symmetric case is that
a general definition of gravitational energy
should give the correct results in various limits:
the asymptotically-flat, vacuum, small-sphere, Newtonian,
test-particle and special-relativistic limits.
In particular,
one would like to obtain the Newtonian mass to first order in $c$,
with corrections interpretable as Newtonian energies.
This is possible for the Penrose energy in certain cases [52].
This brings the discussion back to
the physical meaning of gravitational energy as expressed in the Introduction:
an effective energy which is measurable on a surface
and which is produced by the non-local, non-linear interaction
of the mass of sources with the energy of the consequent gravitational field.
\np\ni
{\bf Acknowledgements.}
It is a pleasure to thank Alan Rendall, Yoshi Fujiwara,
Marcus Kriele, Gregory Burnett and an anonymous referee
for discussions or correspondence which led to improvements in the article.
This work was supported by the Japan Society for the Promotion of Science.
\bs\ni
{\bf Appendix A: Newtonian gravitational energy}
\ms\ni
Recall the dynamical role
of the kinetic energy $K$ and potential energy $V$ (37) in Newtonian gravity,
taking the case of dust, $p=0$.
The Lagrangian $L=K-V$ determines the Euler-Lagrange equation
$$0={\d\over{\d\tau}}\left({\p L\over{\p\dot r}}\right)-{\p L\over{\p r}}
=\I\rho\left(\ddot r+{M\over{r^2}}\right)\eqno(A1)$$
regarding $M$ (or equivalently $*\rho$) as an independent variable
which is conserved, $\dot M=0$.
This yields Newton's inverse-square law of gravitation:\footnote\ddag
{An inverse-square law also exists in the Einstein theory for dust,
with $M$ replaced by $E$ in (A2).
This reinforces the interpretation of $E$ as an active energy.
In this case, $\dot M=\dot E=0$,
so that Lagrangian and Hamiltonian formulations are possible
with $M$ replaced by $E$ in the definition of $V$.}
$$\ddot r=-{M\over{r^2}}.\eqno(A2)$$
The Hamiltonian $H=K+V$ is therefore conserved:
$$\dot H=\I\rho\dot r\left(\ddot r+{M\over{r^2}}\right)=0.\eqno(A3)$$
It is remarkable that both conserved quantities
with a clear physical meaning in Newtonian gravity,
namely the mass $M$ and the Hamiltonian energy $H$,
are encoded in $E$ via the expansion $E=Mc^2+H+O(c^{-2})$ of Proposition 10.

There are two other expressions which are each sometimes suggested as
the potential energy of the Newtonian gravitational field, namely
$$\eqalignno
{&F=-{1\over{8\pi}}\I|\nabla\Phi|^2&(A4a)\cr
&W=\half\I\Phi\rho&(A4b)\cr}$$
where $\Phi$ is the Newtonian potential.
However, neither $F$ nor $W$ coincide with the potential energy $V$
except at infinity, as follows.
Integrating $V$ by parts yields
$$V=\O{-{M^2\over{2r}}}-{1\over{8\pi}}\I{M^2\over{r^4}}.\eqno(A5)$$
Similarly, rewriting $W$ using the Poisson equation
$$\nabla^2\Phi=4\pi\rho\eqno(A6)$$
and integrating by parts yields
$$W=\O{\half r^2\Phi\nabla\!_n\Phi}
-{1\over{8\pi}}\I|\nabla\Phi|^2\eqno(A7)$$
where $n$ is the outgoing unit vector.
The Poisson equation integrates to $\nabla\Phi=nM/r^2$, so that
$$V+\O{{M^2\over{2r}}}=F=W-\O{\half M\Phi}.\eqno(A8)$$
So $V$, $F$ and $W$ all differ by boundary terms.
In particular, at finite radius, $V$ and $F$ differ unless $M$ vanishes,
while $V$ and $W$ differ unless $M$ is constant.
However, if the total mass $\r M$ is finite
and the arbitrary constant in $\Phi$ is fixed as usual by $\r\Phi=0$,
then all three quantities coincide at infinity:
$$\r V=\r F=\r W.\eqno(A9)$$
Thus all three expressions give the correct asymptotic energy.
Nevertheless, at a finite radius, only $V$ gives the correct potential energy
in the Lagrangian or Hamiltonian sense.
Thus it is $V$ rather than $F$ or $W$
that should occur in the Newtonian limit of a quasi-local energy,
even though each would suffice for the asymptotic energy.
\bs\ni
{\bf Appendix B: conserved currents and charges}
\ms\ni
Kodama [53] showed that spherically symmetric space-times
have a conserved current and defined an energy as the associated charge.
This energy is the Misner-Sharp energy, as follows.
Kodama introduced vectors $k$ and $j$ which may be defined,
up to orientation, by
$$k=\hbox{curl}\,r\eqno(B1)$$
and
$$g(j)=-T(k)\eqno(B2)$$
where the curl refers to the two-dimensional quotient space-time.
It follows that
$$g(k,k)={2E\over{r}}-1\eqno(B3)$$
so that $k$ reduces in vacuo (Schwarzschild) to the stationary Killing vector.
One may interpret $k$ as giving a preferred time
for the preferred class of constant-radius observers,
and $j$ as the corresponding energy-momentum density.
Note that $k$ is spatial in trapped regions and causal otherwise.
If the dominant energy condition holds,
$j$ is also causal (or zero) in untrapped regions.
Fixing the orientation so that
$k$ and $j$ are future-causal for untrapped spheres,
whose orientation is chosen as in the main text, one finds explicitly that
$$\eqalignno
{&k=\e^f(\p_+r\p_--\p_-r\p_+)=\e^{-\lambda/2}
\left(r'{\p\over{\p\tau}}-\dot r{\p\over{\p\zeta}}\right)&(B4a)\cr
&j={\e^f\over{4\pi r^2}}(\p_+E\p_--\p_-E\p_+)={\e^{-\lambda/2}\over{4\pi r^2}}
\left(E'{\p\over{\p\tau}}-\dot E{\p\over{\p\zeta}}\right).&(B4b)\cr}$$
Thus $k$ is tangent to the constant-$r$ hypersurfaces
and $j$ is tangent to the constant-$E$ hypersurfaces:
$$\eqalignno
{&k(r)=0&(B5a)\cr
&j(E)=0.&(B5b)\cr}$$
One also has
$$4\pi r^2j=\hbox{curl}\,E.\eqno(B6)$$
Since $\hbox{div}\,\hbox{curl}=0$
and the four-dimensional divergence is given by
$\hbox{Div}\,u=r^{-2}\hbox{div}(r^2u)$,
it follows from (B1) and (B6) that
both $k$ and $j$ are covariantly conserved:
$$\eqalignno
{&\hbox{Div}\,k=0&(B7a)\cr
&\hbox{Div}\,j=0.&(B7b)\cr}$$
By Gauss' law, any such conserved current $u$ has a charge $Q_u=-\I g(u,v)$
which is independent of $\Sigma$ (with fixed boundaries).
Here the notation is that of Section IV, with unit normal vector $v=\p/\p\tau$.
Using (34) and (B4), the charges associated with $k$ and $j$
are the areal volume and energy respectively:
$$\eqalignno
{&Q_k=-\I g(k,v)=\int_0^\zeta4\pi r^2r'\d\hat\zeta=\ft\pi r^3&(B8a)\cr
&Q_j=-\I g(j,v)=\I T(k,v)=\int_0^\zeta E'\d\hat\zeta=E.&(B8b)\cr}$$
The charge $Q_j$ is the definition of energy of Kodama [53].
Regarding $k$ as generating preferred diffeomorphisms,
one may interpret $j$ as the corresponding Noether current
and $E$ as the Noether charge [54].
\bs\ni
{\bf Appendix C: energy-momentum}
\ms\ni
In an asymptotically flat space-time,
there are definitions not just of total energy,
but of total energy-momentum [1,11,26,27],
a vector (or covector) at infinity.
Problems with this interpretation as energy-momentum are described below.
Since this interpretation is often explained by analogy with special relativity,
that case is described first.

In special relativity,
a test particle with mass $m$ and unit velocity vector $u$
has energy-momentum $p=mu$.
In the frame of an observer with unit velocity $v$,
the energy is $-g(p,v)$ and the momentum is $p+g(p,v)v$.
Similarly, a test field with density $\rho$ and velocity $u$
has energy-momentum density $j=\rho u$,
which to an observer has energy density $-g(j,v)$
and momentum density $j+g(j,v)v$.
On a flat spatial hypersurface $\Sigma$ with unit normal $v$,
the total energy $E$, momentum $\Pi$ and energy-momentum $J$ are
$$\eqalignno
{&E=-\I g(j,v)&(C1a)\cr
&\Pi=\I(j+g(j,v)v)&(C1b)\cr
&J=\I j=Ev+\Pi&(C1c)\cr}$$
and the total mass is $\sqrt{-g(J,J)}$.

In general relativity, vectors are not integrable,
so $\Pi$ and $J$ are ill defined.
In contrast, $E$ is well defined for given $j$,
but generally depends on the hypersurface $\Sigma$.
However, it will be independent of $\Sigma$ (with fixed boundaries)
if $j$ is covariantly conserved, by Gauss' law.
As explained in Appendix B,
spherical symmetry admits such an energy-momentum density:
the Kodama current $j$,
a preferred contraction (B2) of the stress-energy tensor.
The corresponding energy $E$ agrees (B8b) with the Misner-Sharp energy.
Again, this identifies $E$ as energy rather than mass.
The total momentum $\Pi$ and energy-momentum $J$ remain ill defined.
(One could make sense of them
by taking the unit spatial vector outside the integral,
but the result would depend on $\Sigma$).
So proceeding by analogy with special relativity yields,
independent of hypersurface,
a total energy but no total momentum, energy-momentum or mass.

Another useful asymptotic quantity is
$$K=\r k\eqno(C2)$$
where $k$ is the Kodama vector (B1).
By (B3), $K$ is a unit vector,
so it may be interpreted as the preferred asymptotic velocity,
or the total velocity of the source.
This provides a preferred frame at infinity,
the asymptotically stationary frame.

Now suppose that one believed that $E$ was mass rather than energy.
Then by analogy with test particles in special relativity, one might define
$$P=\r Ek\eqno(C3)$$
and call it asymptotic energy-momentum.
This applies to both null and spatial infinity,
agreeing with the Bondi-Sachs [11,26,27] and Arnowitt-Deser-Misner [1]
energy-momentum respectively:
$P=E_{BS}K$ and $P=E_{ADM}K$ respectively, by Proposition 3.
Pursuing this idea, one might refer to $-g(P,v)$ and $P+g(P,v)v$
as energy and momentum respectively, and to $E_{BS}$ or $E_{ADM}$ as mass.
This terminology will be avoided here because $E$ is energy rather than mass,
according to the Newtonian, test-particle and special-relativistic limits
of Sections IV and V,
as well as the above analogy with test fields in special relativity.
(If necessary for clarity,
$E_{BS}$ or $E_{ADM}$ may be referred to as the scalar asymptotic energy).
By the same reasoning, $P$ is the product of velocity and energy
rather than velocity and mass, so is not energy-momentum.

To measure the gravitational field experimentally,
one may observe the motion of test particles.
The equation of motion (49) shows that the measurable energy is just $E$,
not something that agrees asymptotically with $-g(P,v)$.

If $P$ did measure asymptotic energy-momentum,
one might expect it to be the limit of a quasi-local energy-momentum,
as for the energy $E$ itself.
But $Ek$ is not recognisable as an energy-momentum at finite radius.
Independently, the energy $E$ and current $k$ are both useful quantities
with simple physical interpretations,
but their product has no obvious physical meaning.
Asymptotically one has the limits $\r E$ and $\r k$,
but their product $P$ also has no obvious physical meaning.
So it seems that there is no asymptotic (or quasi-local) energy-momentum.
Instead one has the radius $r$, the current $k$,
the local energy-momentum density $j$ and the quasi-local energy $E$.
\bs
\begingroup
\parindent=0pt\everypar={\global\hangindent=20pt\hangafter=1}\par
{\bf References}\ms
\refb{[1] Arnowitt R, Deser S \& Misner C W 1962 in}
{Gravitation, an Introduction to Current Research}
{ed: Witten~L (New York: Wiley)}
\ref{[2] Bondi H, van der Burg M G J \& Metzner A W K 1962}\PRS{A269}{21}
\ref{[3] Sachs R K 1962}\PRS{A270}{103}
\ref{[4] Misner C W \& Sharp D H 1964}\PR{136}{B571}
\ref{[5] Penrose R 1973}\ANY{224}{125}
\ref{[6] Hayward S A 1994}\PR{D49}{6467}
\ref{[7] Hayward S A 1994}\CQG{11}{3025}
\ref{[8] Hayward S A 1994}\CQG{11}{3037}
\refb{[9] Hayward S A 1994}{Confinement by black holes}{(gr-qc/9405055)}
\refb{[10] Hawking S W \& Ellis G F R 1973}
{The Large Scale Structure of Space-Time}{(Cambridge University Press)}
\refb{[11] Penrose R \& Rindler W 1986 \& 1988}
{Spinors and Space-Time Volumes 1 \& 2}{(Cambridge University Press)}
\ref{[12] Bizon P, Malec M \& O'Murchadha N 1988}\PRL{61}{1147}
\ref{[13] Penrose R 1969}{Riv.\ Nuovo Cimento}1{251}
\refb{[14] Penrose R 1979 in}{General Relativity, an Einstein Centenary Survey}
{ed: Hawking~S~W \& Israel~W (Cambridge University Press)}
\ref{[15] Hayward S A 1994}\PR{D49}{831}
\refb{[16] Koike T, Onozawa H \& Siino M 1993}
{Causal feature of central singularity and gravitational mass}{(gr-qc/9312012)}
\ref{[17] Christodoulou D 1993}\CPAM{46}{1131}
\ref{[18] Hiscock W A, Williams L G \& Eardley D M 1982}\PR{D26}{751}
\ref{[19] Waugh B \& Lake K 1986}\PR{D34}{2978}
\ref{[20] Eardley D M \& Smarr L 1979}\PR{D19}{2239}
\ref{[21] Christodoulou D 1984}\CMP{93}{171}
\ref{[22] Newman R P A C 1986}\CQG3{527}
\ref{[23] Joshi P S \& Dwivedi I H 1993}\PR{D47}{5357}
\ref{[24] Lake K 1992}\PRL{68}{3129}
\refb{[25] Chiba T, Nakao K \& Nakamura T 1994}
{Masses of spindle and cylindrical naked singularities}{(KUNS-1250)}
\ref{[26] Penrose R 1963}\PRL{10}{66}
\refb{[27] Newman E T \& Tod K P 1980 in}{General Relativity and Gravitation}
{ed: Held~A (New York: Plenum)}
\ref{[28] Ashtekar A \& Hansen R O 1978}\JMP{19}{1542}
\ref{[29] Burnett G A 1991}\PR{D43}{1143}
\ref{[30] Burnett G A 1993}\PR{D48}{5688}
\ref{[31] Malec E \& \'O Murchadha N 1994}\PR{D49}{6931}
\ref{[32] Poisson E \& Israel W 1989}\PRL{63}{1663}
\ref{[33] Poisson E \& Israel W 1989}\PL{B233}{74}
\ref{[34] Poisson E \& Israel W 1990}\PR{D41}{1796}
\ref{[35] M\"uller zum Hagen H, Yodzis P \& Seifert H-J 1974}\CMP{37}{29}
\ref{[36] Kriele M 1993}\CQG{10}{1525}
\refb{[37] Misner C W, Thorne K S \& Wheeler J A 1973}{Gravitation}
{(New York: Freeman)}
\ref{[38] Rendall A D 1994}\CMP{163}{89}
\ref{[39] H\'aj\'\i\v cek P 1987}\PR{D36}{1065}
\ref{[40] Penrose R 1982}\PRS{A381}{53}
\ref{[41] Tod K P 1983}\PRS{A388}{457}
\ref{[42] Dougan A J \& Mason L J 1991}\PRL{67}{2119}
\ref{[43] Bergqvist G 1992}\CQG9{1753}
\ref{[44] Bergqvist G 1992}\CQG9{1917}
\ref{[45] Hawking S W 1968}\JMP9{598}
\ref{[46] Bergqvist G 1994}\CQG{11}{3013}
\ref{[47] Nakamura K, Konno S, Oshiro Y \& Tomimatsu A 1993}\PTP{90}{861}
\ref{[48] Kucha\v r K V 1994}\PR{D50}{3961}
\ref{[49] Brown J D \& York J W 1993}\PR{D47}{1407}
\ref{[50] Bartnik R 1989}\PRL{62}{2346}
\ref{[51] Horowitz G T \& Schmidt B G 1982}\PRS{A381}{215}
\ref{[52] Jeffryes B P 1986}\CQG3{841}
\ref{[53] Kodama H 1980}\PTP{63}{1217}
\refb{[54] Fujiwara Y 1995}
{Local energy in spherical space-times as Noether charge}{(YITP/U-95-06)}
\endgroup
\bye